\documentclass[pre,twocolumn,superscriptaddress,amsmath,amssymb,floatfix,showpacs]{revtex4}
\usepackage{graphicx}
\usepackage{epsfig}
\usepackage{dcolumn}
\usepackage{bm}
\usepackage{psfrag}
\usepackage{gensymb}

\newcommand{\Evec}{\mathbf{E}}

\newcommand{\nvec}{\mathbf{n}}

\newcommand{\dd}{\mathrm{d}}

\begin{document}

\bibliographystyle{apsrev}

\title{Nematic droplets at fibres}

\author{V. M. O. Batista}
\affiliation{Departamento de F{\'\i}sica da Faculdade de Ci\^encias,}
\affiliation{Centro de F{\'\i}sica Te\'orica e Computacional,
Universidade de Lisboa, Campo Grande, P-1649-003 Lisboa, Portugal}

\author{N. M. Silvestre}
\email[]{nmsilvestre@fc.ul.pt}
\affiliation{Departamento de F{\'\i}sica da Faculdade de Ci\^encias,}
\affiliation{Centro de F{\'\i}sica Te\'orica e Computacional,
Universidade de Lisboa, Campo Grande, P-1649-003 Lisboa, Portugal}

\author{M. M. Telo da Gama}
\affiliation{Departamento de F{\'\i}sica da Faculdade de Ci\^encias,}
\affiliation{Centro de F{\'\i}sica Te\'orica e Computacional,
Universidade de Lisboa, Campo Grande, P-1649-003 Lisboa, Portugal}

\date{\today}

\begin{abstract}
The emergence of new techniques for the fabrication of nematic droplets with nontrivial topology provides new routes for the assembly of responsive devices. 
Here we perform a numerical study of spherical nematic droplets on fibres. We analyse the equilibrium textures and find that, under certain conditions, 
the nematic can avoid the nucleation of topological defects. We consider in detail a homeotropic nematic droplet wrapped around a fibre with planar anchoring. 
We investigate the effect of an electric field on the texture of this droplet such type of system. In the presence of a DC field, the system undergoes a Freederickzs-like 
transition above a given threshold $E_c$. We also consider AC fields, at high and low frequencies, and find that the textures are similar to those observed for static fields, 
in contrast with recently reported experiments.

\end{abstract}

\pacs{...}

\maketitle

\section{Introduction}

In the last decade, the structure and dynamics of nematic droplets have been studied, theoretically and experimentally, due to their key role in novel 
applications based on polymer dispersed liquid crystals (PDLCs) \cite{Drzaic.2006}. 
The interplay between the spherical confinement and the molecular alignment at the surface of the droplets leads to the nucleation of topological defects. 
In general, two types of surface alignment (anchoring) are considered: homeotropic (perpendicular) and planar (parallel). Nematic liquid crystals confined to 
homeotropic spherical drops can exhibit a radial configuration with a point-like defect, of topological charge $+1$, at the center of the drop. The radial configuration 
is observed in large droplets, with radii $R>3\mu$m \cite{Ondris-Crawford.1991}. For smaller droplets, however, the nematic texture adopts what is known as an axial 
configuration with a topological defect ring of cross-sectional winding number $+1/2$. The competition between the splay energy and the anchoring energy makes the point 
defect energetically unfavourable against the ring for sufficiently small drops \cite{Mori.1988,Erdmann.1990}. This transition may also be driven by applying an 
electric field \cite{Ondris-Crawford.1991} resulting in distinct optical responses as the field is switched on or off. 

On the other hand, if the surface of the nematic drop induces parallel alignment, the droplet exhibits a bipolar configuration with two surface defects (boojums) nucleated 
at antipodal points. In ellipsoidal droplets the defects are forced to the regions of higher curvature to relieve splay and bend elastic stress \cite{Volovik.1983}. For 
some liquid crystals, such as 8CB, planar nematic droplets with a bipolar configuration can undergo a twist transition where the nematic director twists from one surface 
defect to the other. This transition is controlled by temperature, which turns twist deformations energetically favourable against bend and splay deformations \cite{Lavrentovich.1990}. 
The twist transition does not affect the type of defects nucleated at the droplet surface. However, in systems where it is energetically preferable to have bend 
deformations rather than splay deformations, the nematic takes a concentric configuration with a defect line connecting the poles of the droplet along the symmetry axis \cite{Drzaic.1988}. 
Planar nematic droplets with bipolar configurations are the most common. Of particular interest are those with prolate shapes that allow the switching between two 
distinct nematic configurations with the application and removal of an external electric field \cite{Drzaic.2006}.

With the development of new soft-lithographic techniques it is now possible to produce in a controlled fashion new PDLC matrices with nematic droplets of prescribed shapes 
and non-trivial topologies. This gave origin to what was coined as topological PDLCs (TPDLCs) \cite{Campbell.2014}. Of particular interest are toroidal shapes, 
for which the Hopf-Poincar\'e theorem dictates that the net charge of the defects is $\sum_i{q_i}=0$, for homogeneous surface alignment. It turns out that for a 
torus on the micrometer scale defects may be present in pairs of opposite charge \cite{Campbell.2014}. However, as the system size increases the presence of topological 
defects becomes energetically unstable, and for a torus on the millimeter scale, the nematic texture is completely free from topological defects \cite{Pairam.2013}.

A topology which is also toroidal is that of a droplet adsorbed around a microfibre \cite{Geng.2012,Geng.2013}. In this case, however, the system has two confining surfaces 
that can differ in anchoring: ($i$) that of the droplet itself and ($ii$) the surface of the microfibre. Fabrication of these systems is easily achieved but control of the 
size of the droplets is limited. They were reported in \cite{Geng.2012}, where 5CB liquid crystal droplets were deposited on $1 \,\mu$m thick fibres through aerosol evaporation. 
The droplets, with diameters between $10$ to $30$ $\mu$m, are in contact with air and the nematic texture adopts the axial configuration, with a ring defect around the fibre, 
indicating that the two surfaces have antagonistic anchorings. At the nematic-air interface the 5CB molecules are aligned perpendicular to the interface, implying that the surface 
of the fibre imposes planar (or planar degenerate) alignment. The advantage of this type of systems arises from the possibility of changing the surface alignment through 
the adsorption of certain molecules. For example, endotoxins are known to trigger ordering transitions in nematic droplets from the bipolar (planar alignment at the surface) to 
the radial (perpendicular alignment) configuration \cite{Miller.2013}. Also, the defect configuration may be strongly deformed with applied electric fields due to the presence of 
the fibre, which pierces the droplet. The application of electric fields may change the configuration of the defect in isolated droplets but in most cases a rotation of the nematic 
texture has been reported \cite{Doane.1988}.

Here we investigate the equilibrium textures of nematic droplets adsorbed at fibres for different boundary conditions. We show that the nucleation of topological defects in the LC 
matrix depends on the matching (or mismatching) anchoring conditions at the fibre and surface of the droplet, and it may be avoided by allowing the nematic molecules 
to twist around the fibre. We will focus in particular on the case of homeotropic droplets pierced by a planar fibre and investigate how the system responds to 
applied external electric fields.

\section{Model}
\label{sec:model}

The nematic order of the fluid is expressed using a traceless, symmetric, tensorial order parameter~\citep{degennes93} called the Q-tensor,
\begin{equation}
Q_{\alpha\beta}=\tfrac{1}{2}S\left(3n_{\alpha}n_{\beta}-\delta_{\alpha\beta}\right)+\tfrac{1}{2}B\left(m_{\alpha}m_{\beta}-l_{\alpha}l_{\beta}\right),   \label{eqn:QTensor}
\end{equation}

\noindent where $\mathbf{n}$ is a unit vector denoting the director, and $S$ is the degree of nematic ordering. 
In some circumstances there may be biaxial ordering of 
degree $B$, with $\mathbf{m}$ and $\mathbf{l}$ forming an orthonormal set with $\mathbf{n}$.

The free energy of the system is given by the functional over the fluid region $\mathcal{R}$ and the boundaries $\mathcal{W}$
\begin{multline}
\mathcal{F}=\int_{\mathcal{R}}\Big\{\tfrac{2}{3}A\left(\tau^* S_{\text{nem}}^{-2}Q_{\alpha\beta}Q_{\beta\alpha} \right. \\
\left. -\tfrac{4}{3}(2+\tau^*)S_{\text{nem}}^{-3}Q_{\alpha\beta}Q_{\beta\gamma}Q_{\gamma\alpha} \right.
\left. +\tfrac{2}{3}S_{\text{nem}}^{-4}\left[Q_{\alpha\beta}Q_{\beta\alpha}\right]^{2}\right) \\
\left.  + \tfrac{1}{2}L_{1}\partial_{\gamma}Q_{\alpha\beta}\partial_{\gamma}Q_{\alpha\beta} + \right.
\left.  + \tfrac{1}{2}L_{2}\partial_{\alpha}Q_{\alpha\gamma}\partial_{\beta}Q_{\gamma\beta}\Big\}\dd V + \right. \\
\int_{\mathcal{W}}\tfrac{1}{2}\alpha(Q_{\alpha\beta}^{\text{pref}}-Q_{\alpha\beta})^{2}\dd S,    
\label{eqn:freeEnergy}
\end{multline}
\noindent where $A$, $L_{1}$ and $L_{2}$, and $\alpha$ are positive coefficients for bulk, elastic and anchoring free energies respectively. $\tau^*$ is a reduced temperature, 
such that the nematic phase with $S=S_{\text{nem}}$ is favoured when $\tau^*<1$, and the isotropic (unordered) phase with $S=0$ is favoured 
when $\tau^*>1$. $Q_{\alpha\beta}^{\text{pref}}$ is the value of the Q-tensor preferred by the anchoring at the boundary. In all cases reported here we choose the elastic 
constants such that $L_{2}=2L_{1}$. This corresponds to liquid crystal compounds with a twist elastic constant lower than splay and bend elastic constants, as 5CB.

Within $\mathcal{R}$, the fluid density $\rho$, velocity $u$, and Q-tensor evolve over time $t$ according to the continuity, Navier-Stokes, and Beris-Edwards~\citep{Beris94} equations:

\begin{equation}
\partial_{t}\rho+\partial_{\beta}(\rho u_{\beta})=0      
\label{eqn:continuity}
\end{equation}
\begin{multline}
\rho\left(\partial_{t}+u_{\beta}\partial_{\beta}\right)u_{\alpha}=\\
\partial_{\beta}\left[2\mu\Lambda_{\alpha\beta}-p\delta_{\alpha\beta}+\left\{\zeta\Sigma_{\alpha\beta\gamma\delta}+ 
 \mathrm{T}_{\alpha\beta\gamma\delta}\right\}H_{\gamma\delta}\right]\\
 -H_{\beta\gamma}\partial_{\alpha}Q_{\gamma\beta}
\label{eqn:navierStokes} 
\end{multline}
\begin{equation}
\left(\partial_{t}+u_{\gamma}\partial_{\gamma}\right)Q_{\alpha\beta}=
-\zeta\Sigma_{\alpha\beta\gamma\delta}\Lambda_{\gamma\delta}-\mathrm{T}_{\alpha\beta\gamma\delta}\Omega_{\gamma\delta}+\Gamma H_{\alpha\beta} 
\label{eqn:berisEdwards}
\end{equation}

with

\begin{equation}
\begin{split}
H_{\alpha\beta}&=\frac{1}{3}\frac{\delta \mathcal{F}}{\delta Q_{\gamma\gamma}}\delta_{\alpha\beta}-\frac{1}{2}\left(\frac{\delta \mathcal{F}}{\delta Q_{\alpha\beta}}+\frac{\delta \mathcal{F}}{\delta Q_{\beta\alpha}}\right),\\
\Sigma_{\alpha\beta\gamma\delta}&=\tfrac{4}{3}S_{\text{nem}}^{-1}Q_{\alpha\beta}Q_{\gamma\delta}-\delta_{\alpha\gamma}(Q_{\delta\beta}+\tfrac{1}{2}S_{\text{nem}}\delta_{\delta\beta})\\
& -(Q_{\alpha\delta}+\tfrac{1}{2}S_{\text{nem}}\delta_{\alpha\delta})\delta_{\gamma\beta}+\tfrac{2}{3}\delta_{\alpha\beta}(Q_{\gamma\delta}+\tfrac{1}{2}S_{\text{nem}}\delta_{\gamma\delta}),\\
\mathrm{T}_{\alpha\beta\gamma\delta}&=Q_{\alpha\gamma}\delta_{\beta\delta}-\delta_{\alpha\gamma}Q_{\beta\delta},\\
\Lambda_{\alpha\beta}&=\tfrac{1}{2}\left(\partial_{\beta}u_{\alpha}+\partial_{\alpha}u_{\beta}\right),\\
\Omega_{\alpha\beta}&=\tfrac{1}{2}\left(\partial_{\beta}u_{\alpha}-\partial_{\alpha}u_{\beta}\right),
\end{split}
\end{equation}

\noindent where $p=\rho/3$ is the isotropic fluid pressure, $\mu$ is the dynamic viscosity, and $\Gamma$ is the mobility of the nematic order. $\zeta$ is a dynamical parameter, 
dependent on the molecular details of the liquid crystal, which determines how the nematic orientation couples to shear. 
If $\zeta<1$, then the director will tumble indefinitely in the shear, while if $\zeta>1$ (which we shall consider 
in this paper) permits a bulk, state-state orientation of the director relative to the shear \citep{leslie68}.

At the boundaries  $\mathcal{W}$, non-slip and anchoring conditions apply,
\begin{align}
(\delta_{\alpha\beta}-\nu_{\alpha}\nu_{\beta})u_{\beta}&=0, \label{eqn:nonslip} \\
L_1\nu_{\gamma}\partial_{\gamma}Q_{\alpha\beta} + L_2 \nu_\alpha\partial_\gamma Q_{\beta\gamma}&=\alpha\left(Q_{\alpha\beta}^{\text{pref}}-Q_{\alpha\beta}\right),  \label{eqn:substrateEquilibrium}
\end{align}
where $\pmb{\nu}$ is the inward normal to the substrate. As an alternative to Eqn.~\ref{eqn:substrateEquilibrium}, we may impose the boundary 
condition $\mathbf{Q}=\mathbf{Q}^{\text{pref}}$. This is equivalent to setting $\alpha=\infty$.

\section{Simulation method}
We simulate the dynamics of the fluid by discretising space and time - the former into a cubic grid of nodes - and maintain $\rho$, $\mathbf{u}$ and $\mathbf{Q}$ as continuous quantities. 
We utilise a hybrid method in which Eqns.~\ref{eqn:continuity} and \ref{eqn:navierStokes} are iterated using a lattice Boltzmann method and Eqn.~\ref{eqn:berisEdwards} using a
a finite-difference method~\citep{marenduzzo07}, a method that has been previously used by our group when considering a nematic liquid crystal in contact with a 
substrate patterned with rectangular grooves that may fill without the occurrence of complete wetting \cite{Blow2013}. 
In that study our numerical code was used to analyse the dynamical response of the system to an externally-applied electric field so as to identify switching transitions 
between these filled states and was also applied by our group to study the effect of anchoring strength in the flow of a nematic liquid crystal \cite{Batista2015}.

To incorporate an electric field E into the simulation we take into account the electric contribution to the free energy

\begin{equation}
F_{\rm{elec}}=-\tfrac{1}{2}\epsilon_{0} (\epsilon_{I}\delta_{\alpha\beta}+\epsilon_{A}S^{-1}_{nem}Q_{\alpha\beta}) E_{\alpha}E_{\beta}      
\label{eqn:Felec}
\end{equation}

\noindent where $\epsilon_{0}$ is the electric permittivity of vacuum, $\epsilon_{I}$ the relative permittivity  and $\Delta\epsilon$ the dielectric anisotropy. 
$\Delta\epsilon>0$ corresponds to systems where particles tend to orient along the direction of the electric field, whereas for $\Delta\epsilon<0$ particles 
orient perpendicular to the field. In this study, which is meant to model 5CB, we consider that particles align 
in the direction of the electric field with $\Delta\epsilon=11.5$. An electric field $E$ (AC) is defined as 

\begin{equation}
E(t)=E_{\rm{mag}}\cos(2\pi f t)      
\label{eqn:Eelec}
\end{equation}

\noindent with $E_{\rm{mag}}$ the magnitude of the electric field and $f$ its frequency. Setting $f=0$ reverts to a DC field of magnitude $E_{\rm{mag}}$. 
We follow experimental analyses and also many other numerical studies by assuming uniform $E$ throughout the simulation box.

Simulation parameters must be mapped to physical quantities of 5CB by an appropriate choice of scaling. Using reported data for 5CB, 
$A\sim 10^5 \rm{Nm^{-2}}$, $L_{1}\sim10^{-12} \rm{N}$ with $L_{2}=2L_{1}$, $\xi\sim10^{-8} \rm{m}$, nematic-air surface tension $\gamma=0.038 \rm{Nm^{-1}}$ 
and isotropic-nematic transition between $18$ and $34 \degree$ Celsius, we set our simulation parameters 
to $A=0.188$, $L_{1}=0.085$, $L_{2}=2L_{1}$, $\Gamma=0.25$, $\alpha=1$, 
$\rho=80$, (with deviations due to the effects of compressibility, which are small), $\zeta=1.5$ (aligning regime) and $S_{\text{nem}}=1$.   
Hence the physical length-scale corresponding to a lattice spacing, is $\Delta x \sim 10^{-8} \,\rm{m}$, and the time scale is $\Delta t \sim 10^{-6}\, \rm{s}$. $E_{\rm{mag}}$ 
and $f$ are  parameters of interest that we shall vary.

\section{Nematic droplets adsorbed at fibres}
\subsection{In the absence of E fields}

The texture or equilibrium configuration of a nematic droplet pierced by a fibre depends on (i) the anchoring conditions at both the fibre and the nematic-air interface, 
and (ii) the ratio between the size of the droplet and the diameter of the fibre that pierces it, $d_{drop}/d_{fibre}$. We start by considering different types of anchoring conditions. 
As discussed previously, nematic droplets with either homeotropic (perpendicular) or planar (parallel) alignment at their surface have topological charge of $+1$ which induces the 
nucleation of defects either in bulk or at the surface, respectively. By piercing the droplet with a fibre, the structure of the defects can change or, in some cases, they can disappear. 
This depends on the type of anchoring at the fibre compared to the anchoring at the nematic-air interface.
 
\begin{figure}
\includegraphics[width=0.9\columnwidth]{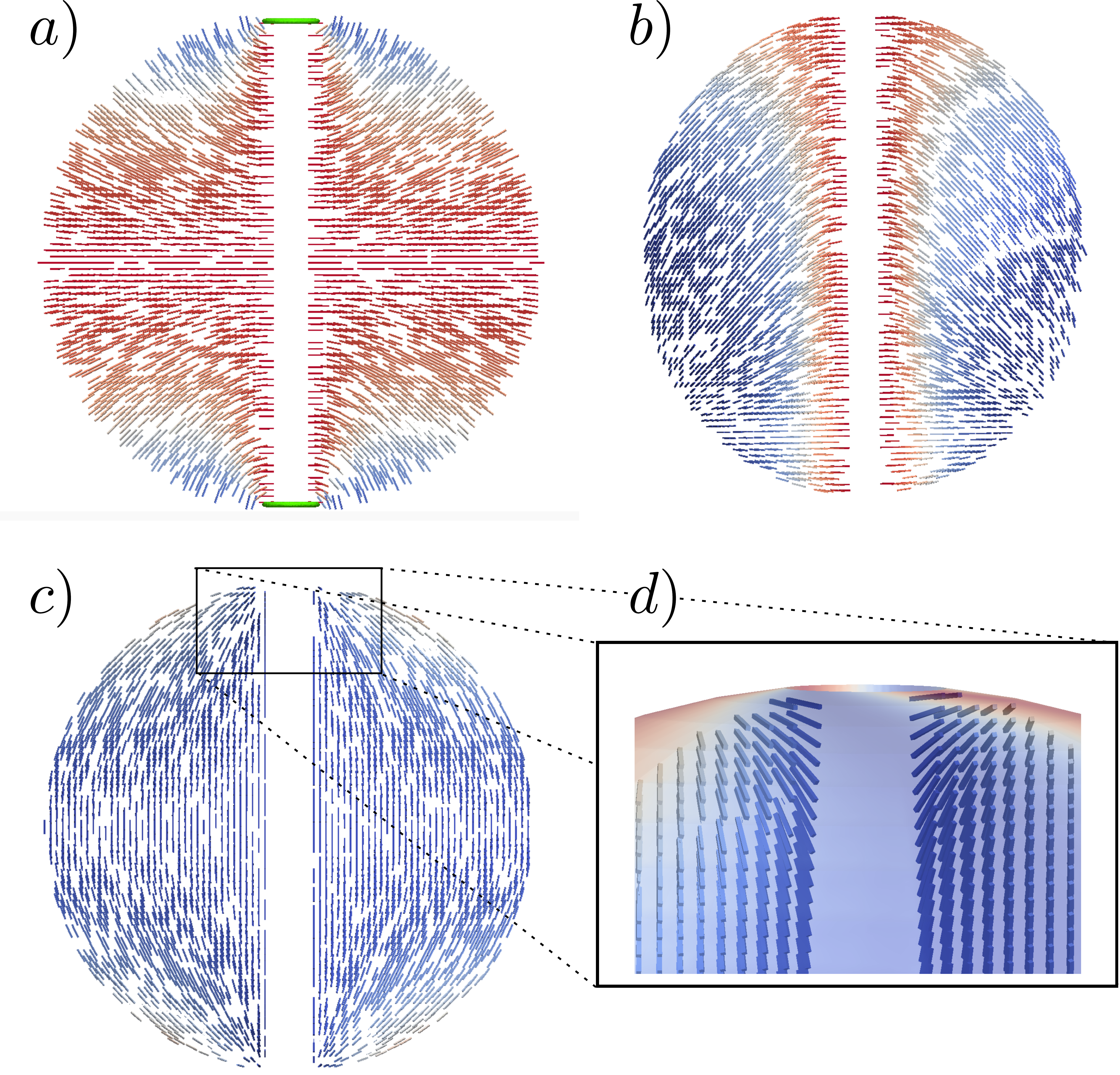}
\caption{(color online) Nematic droplets at fibres with different anchoring conditions. $a)$ Homeotropic alignment at both the fibre and the nematic-air interface; $b)$ homeotropic alignment at the fibre and planar degenerate at the nematic-air interface; $c)$ and $d)$ planar degenerate at both the fibre and the nematic-air interface. The bars represent the director field $\nvec$ and the color code indicates the alignment of $\nvec$ with the orientation perpendicular to the fibre; red if $\nvec$ is perpendicular and blue if $\nvec$ is parallel to the fibre. The topological defect is represented by the isosurface in green.}
\label{diff_drops}
\end{figure} 

\begin{figure}
\includegraphics[width=0.9\columnwidth]{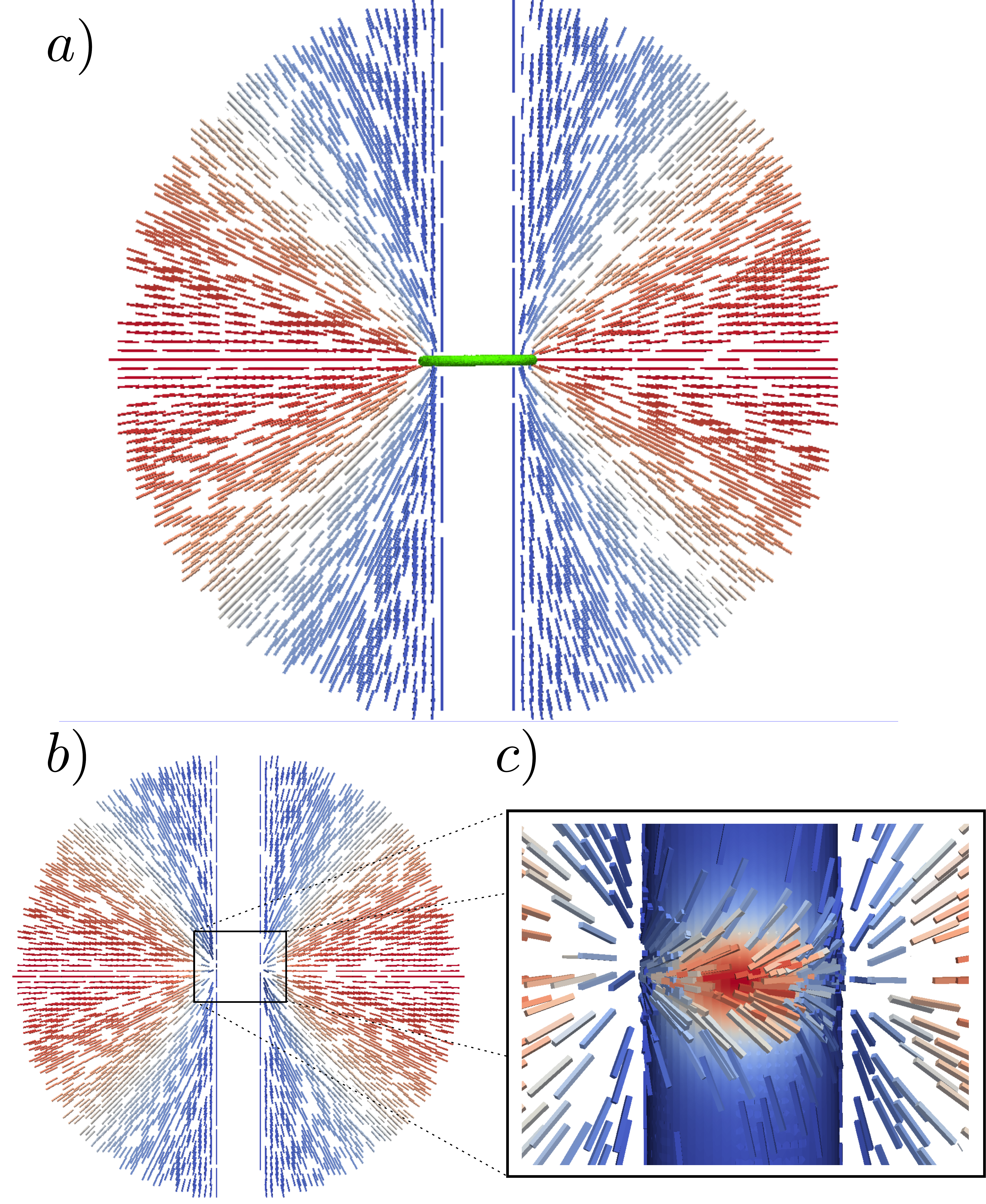}
\caption{(color online) Homeotropic nematic droplet at a planar fibre. $a)$ Equilibrium configuration with a ring defect surrounding the fibre at the middle of the droplet. If the anchoring on the fibre is planar degenerate, $b)$ there is no topological defect in the nematic and $c)$ the liquid crystal texture assumes an escaped configuration close to the fibre. The bars represent the director field $\nvec$ and the color code indicates the alignment of $\nvec$ with the orientation perpendicular to the fibre; red if $\nvec$ is perpendicular and blue if $\nvec$ is parallel to the fibre. The topological defect is represented by the isosurface in green. }
\label{hybrid_drops}
\end{figure}

\begin{figure}[t]
\includegraphics[width=\columnwidth]{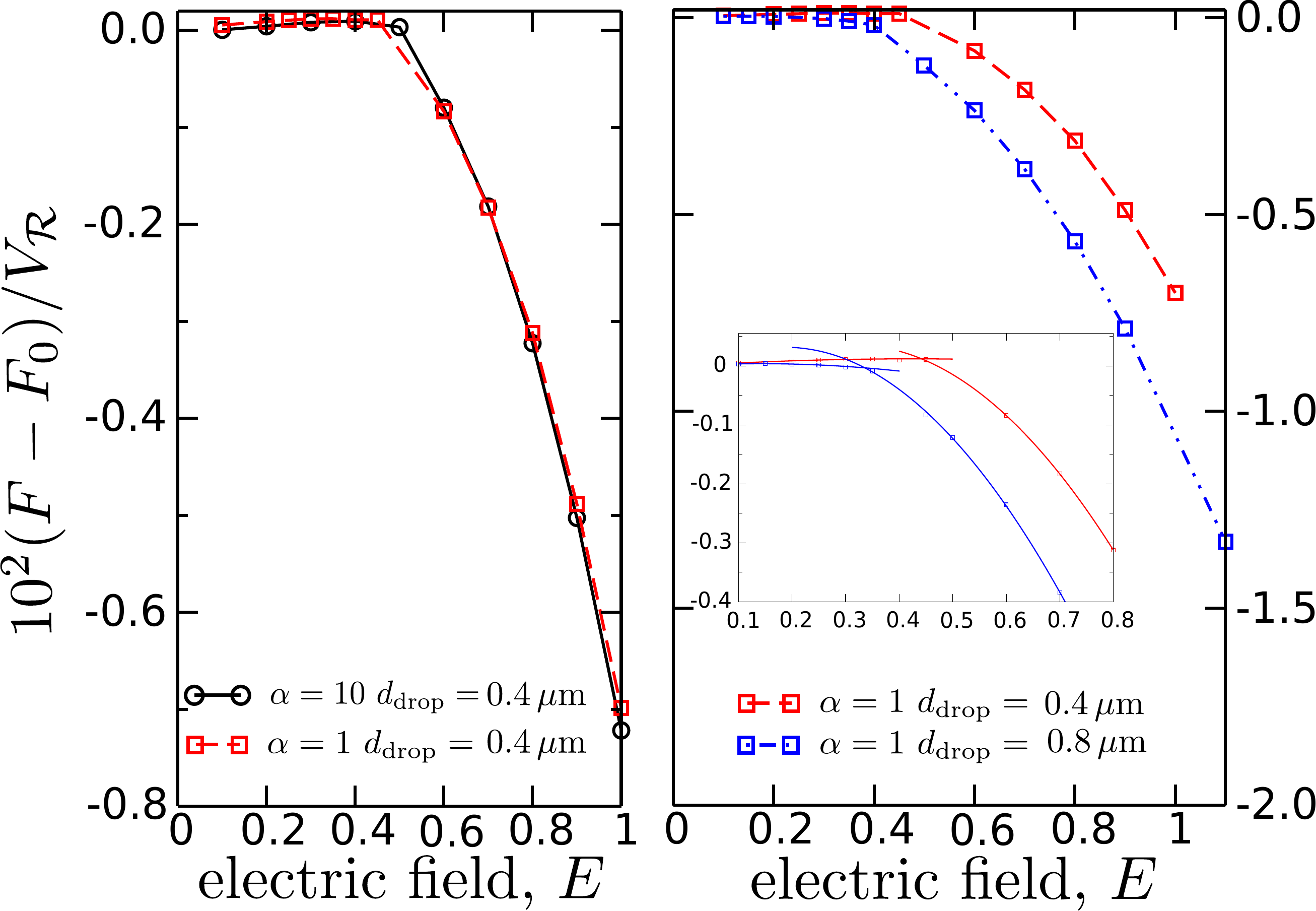}
\caption{(color online) Free energy $F$ of a homeotropic nematic droplet adsorbed at a fibre as a function of the applied DC electric field $E$. $F_0=F(E=0)$ is the free energy in the absence 
of the external field and $V_{\mathcal{R}}$ is the LC drop volume. Left: fixed system size $d_{drop}=0.4\,\mu$m and different anchoring strengths $\alpha=1$ ($\square$, red),$10$ ($\circ$, black). 
Right: fixed anchoring strength $\alpha=1$ and different system sizes $d_{drop}=0.4\,\mu$m (dash line, red), $0.8\,\mu$m (dash-dot line, blue). The inset illustrates the dependence of $E_c$, 
given by the intersection of the same-coloured curves, on system size, as explained in the text.}
\label{Freed}
\end{figure}

\begin{figure*}[t]
\includegraphics[width=1.4\columnwidth]{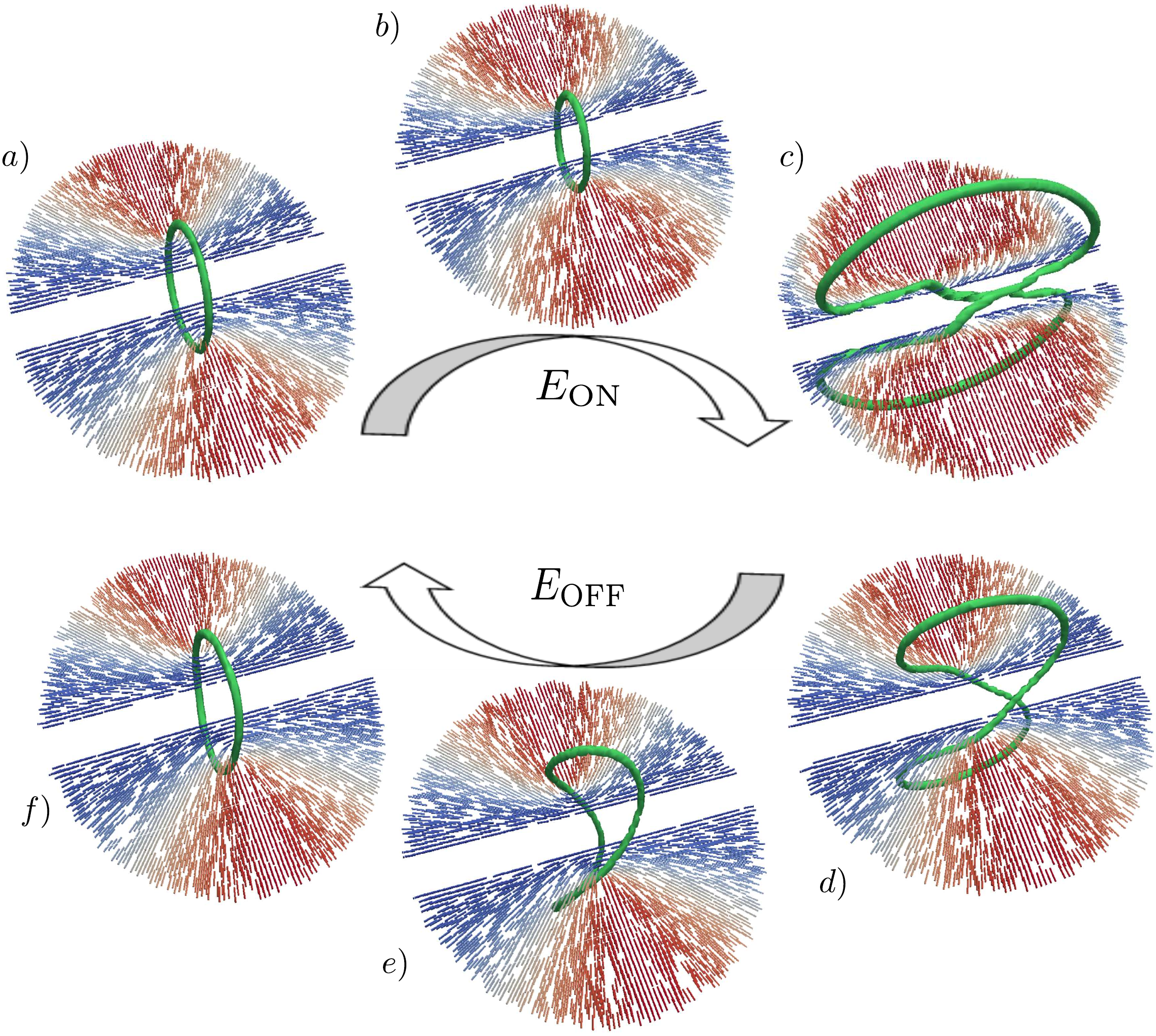}
\caption{(color online) Cross section of the liquid crystal texture of a homeotropic nematic droplet adsorbed at a fibre $a)$ at equilibrium with $E=0$, $b)$ and $c)$ under the influence of a DC electric field for field strengths respectively below and above the Freederickzs transition, and $d)$ to $f)$ evolving towards the equilibrium configuration after switching off the electric field. The snapshots for the off sequence were taken for $d)$ $t=5\,m$s, $e)$ $t=10\,m$s, and $f)$ $t=25\,m$s. The bars represent the director field $\nvec$ and the color code indicates the alignment of $\nvec$ with the applied electric field; red if $\nvec\parallel\Evec$ and blue if $\nvec\perp\Evec$. The topological defect is represented by the isosurface in green. }
\label{Efieldsequence}
\end{figure*}

\begin{figure*}
\includegraphics[width=1.5\columnwidth]{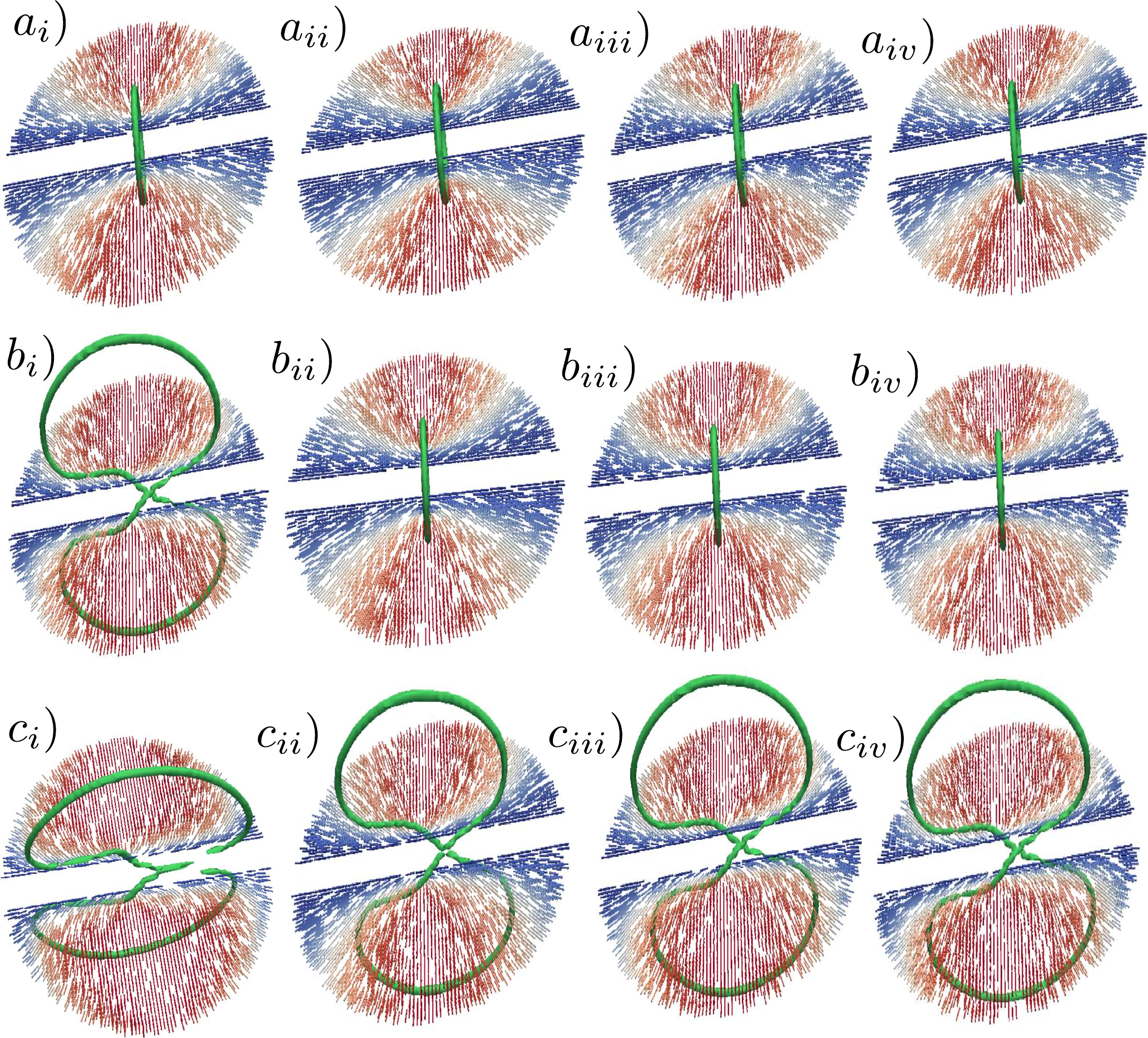}
\caption{(color online) Cross section of the liquid crystal texture of a homeotropic nematic droplet adsorbed at a fibre under the influence of an AC electric field for field strengths $a)$ below the Freederickzs transition, $b)$  close to and above the Freederickzs transition, and $c)$ far and above the Freederickzs transition, and for frequencies $i)$ $f=0$ Hz (DC), $ii)$ $f\sim 100$ Hz, $iii)$ $f=1\,k$Hz, and $iv)$ $f=100\,k$Hz. The bars represent the director field $\nvec$ and the color code indicates the alignment of $\nvec$ with the applied electric field; red if $\nvec\parallel\Evec$ and blue if $\nvec\perp\Evec$. The topological defect is represented by the isosurface in green.}
\label{efreq}
\end{figure*}
 
In Fig.\ref{diff_drops} we show three nematic droplets adsorbed at fibres for different anchoring conditions. When the surface of the droplet induces homeotropic anchoring a defect is 
nucleated in bulk at the center of the drop. As discussed before, this can be either a point defect, which typically appears for large droplets and the nematic has a radial configuration, 
or a ring defect with a nematic texture that is axial. In Fig.~\ref{diff_drops}$a$ a fibre with homeotropic anchoring goes through the center of the nematic drop. The mere presence of a 
fibre going through the centre of the drop makes it physically impossible for the defect to nucleate in that same region. In the case of a homeotropic fibre this is 
compensated by the nucleation of two surface defects at the contact lines (see Fig.\ref{diff_drops} in green), and no defect is nucleated in bulk. Although we did not explore it further, 
as our droplet is assumed to be rigid, in this situation its shape is expected to elongate. Such elongation would be balanced by the surface tension.

Nematic droplets with planar degenerate anchoring can either exhibit a bipolar or concentric texture, as mentioned before. However, in the presence of a homeotropic fibre, the defects of 
either configurations are no longer nucleated. In Fig.\ref{diff_drops}$b$ we see that instead of nucleating topological defects, the nematic assumes an escaped configuration around 
the fibre, where the director field seems to rotate around the fibre. This is possible due to the degenerate orientation at the surface of the droplets.
A similar effect is seen when one considers a fibre with planar (degenerate) anchoring, shown in Figs.\ref{diff_drops}$c$ and $d$. 
In this case the director field rotates close to the contact lines, clearly seen in Fig.\ref{diff_drops}$c$, where the director assumes a perpendicular 
orientation to the fibre (symmetry) axis, while planar to its surface, and as it moves away from the contact line towards the center, the director aligns parallel to the symmetry axis.


Finally, if a fibre with planar anchoring pierces a homeotropic droplet the nematic nucleates a topological ring defect close to the fibre as seen in Fig.\ref{hybrid_drops}$a$. This resembles the axial configuration of nematic drops, where the fibre plays the role of the axially uniform nematic. This configuration is stable for any size of the droplet if the anchoring on the fibre is non-degenerate. However, if the fibre has planar degenerate anchoring the ring defect is unstable for large systems. It turns out that by increasing the effective distance between the surface of the nematic drop and the surface of the fibre, while fixing $d_{drop}/d_{fibre}$, the nematic molecules are able to rotate around the fibre at the center of the drop, thus removing the topological defect (Fig.\ref{hybrid_drops}$b$) and assuming an escaped configuration that induces bend-splay deformations (Fig.\ref{hybrid_drops}$c$). Such configuration is indeed not reported in the experiments \cite{Geng.2012} which hints that the fibre may induce a preferred planar orientation rather than degenerate. For small systems with planar degenerate anchoring on the fibre the escaped configuration is metastable and can be achieved through the application of an external electric field.  We note that if the surface of the nematic drop is contaminated by an agent that induces planar anchoring the liquid crystal texture would switch from a configuration with a ring defect to a configuration without any topological defects, inducing a different optical response.

\subsection{In the presence of DC E fields}

Homeotropic nematic droplets on fibres with planar anchoring are feasible from the experimental point of view. The appearance of a ring defect around the fibre enhances the potential of such systems for applications, as an external field can induce an orientational state with an optical response that is significantly different from that of the ground state. Here we are interested on the response to applied electric fields. For comparison we start by assessing the effect of electric fields produced by direct current (DC) electrostatic potentials. Alternating current (AC) fields will be addressed in the next section.

Here we consider that the electric field $E$ is perpendicular to the fibre and couples with the nematic LC via the free energy density given in Eq.\ref{eqn:Felec}. When applying an electric 
field the system will respond in accordance with the competition between the elastic energy, that acts as a restoring force, and the field energy. If the field is strong enough 
it will deform the orientational ground state and the system will take a characteristic time to reach the new state that is a product of such competing mechanisms. Figure \ref{Freed} shows 
how the free energy $F$ changes as the applied DC electric field $E$ is increased. As expected, the system undergoes a Freederickzs type transition, well known for flat liquid 
crystal cells \cite{degennes93, Chaikin.2000}. Here, although the geometry of the system is more complex, the response to electric fields seems to be very similar. For weak field strengths, 
little reorientation of the nematic director occurs, the elastic strength is strong and the systems remains very close to its ground state configuration. Above a critical field strength, $E_c$, 
the electric field aligning forces are stronger than the elastic ones and the nematic tends to align with the $E$ field, thus deforming completely the zero field state. 
In Fig. \ref{Freed} we show the free energy $F$ as a function of the applied electric field $E$ for (left) fixed system size $d_{drop}=0.4\,\mu$m and different anchoring strengths $\alpha=1,\,10$, 
and (right) fixed anchoring strength $\alpha=1$ and different system sizes $d_{drop}=0.4,\,0.8\mu$m. $F_0$ is the free energy for $E=0$ and $V_{\mathcal{R}}$ is the volume of the nematic liquid crystal. 
For the same system size, increasing the anchoring strength results in an increase of the transition field. For example, for $d_{drop}=0.4\,\mu$m increasing the anchoring from $\alpha=1$ 
to $\alpha=10$ induces the transition field to change from $E_c=0.40$ to $E_c=0.45$. On the other hand, for the same anchoring strength $\alpha=1$, changing the system size 
from $d_{drop}=0.4\mu$m to $d_{drop}=0.8\mu$m has the reverse effect, and the transition field decreases from $E_c=0.40$ to $E_c=0.34$. As the anchoring strength is 
increased the aligning effect of the surfaces gets stronger and it increases the energy cost of uniformly aligning the liquid crystal molecules along the field, 
which results in a higher transition field. Enlarging the separation between the confining surfaces weakens this effect and the transition field is lower.

To address the response of the system to the electric field, we show in Fig. \ref{Efieldsequence} what happens to the nematic configuration when the field is turned up to a 
high field strength and then switched off. When there is no electric field and the system is at equilibrium, the nematic exhibits the axial configuration of Fig.\ref{Efieldsequence}$a$ with 
a defect ring around the fibre. As the electric field is switched on up to a value that is below the Freederickzs transition ($E<E_c$) the LC molecules start to align 
with the field. This process is counteracted by the elastic stress and the system remains in the axial configuration. However, we do observe a contraction of the 
ring defect as an effect of the electric field (Fig.\ref{Efieldsequence}$b$). If the field strength is higher than $E_c$ the nematic molecules mostly align with the field. In a nematic droplet, 
as the contraction of the defect becomes unstable, this would lead to a rotation of the defect ring, which would also expand towards the surface of the drop. Here, given 
the presence of the fibre, the ring rotates (twists) in two different directions while expanding towards the surface of the drop, transforming the ring into a figure-of-eight, 
as represented in Fig.\ref{Efieldsequence}$c$. When switching off the electric field the elastic stress acts in order to restore the system to a state with lower elastic deformations. 
This results in the untwisting of the figure-of-eight defect (Fig.\ref{Efieldsequence}$d$) and the contraction towards the fibre (Fig.\ref{Efieldsequence}$e$ and $f$).

The time it takes for the system to achieve the new state depends on the processes involved. Always present is the elastic stress that acts as a restoring force that takes the system into a 
state of lower elastic deformation. When applying an electric field the response of the system depends not just on the restoring effect of the elastic stresses but also on the 
perturbing/aligning effect of the electric field, and has a characteristic time given by $\tau_{on}=1/\left(\tau_N^{-1} + \tau_E^{-1}\right)$ \cite{Doane.1988}, where $\tau_N$ and $\tau_E$ 
are respectively the characteristic times associated with the elastic and electric stresses.  For very large field strengths $\tau_{on}\sim\tau_E$. When the electric 
field is switched off the relaxation time depends solely on the characteristic time due to the elastic stress, $\tau_{off}\sim\tau_N$. 

The relevant characteristic times can be estimated by the dimensional analysis of Eq. \ref{eqn:berisEdwards}. This indicates that the characteristic time associated with the elastic stress 
scales with $\tau_N\propto \lambda^2/\left(\Gamma L_1\right)$, where $\lambda$ is a characteristic length. For (transient) configurations close to the (equilibrium) axial 
configuration $\lambda\sim d_{fibre}$. In the presence of an electric field $E>E_c$ the configuration is extremely different from the equilibrium one given that the defect 
expands to the surface of the drop. In spherical, or ellipsoidal, nematic drops the defect would expand maintaining its ring shape and in that case $\lambda\sim d_{drop}$. 
In the presence of a piercing fibre this is no longer the case and $\lambda$ is not as trivial to estimate. There are, however, at least two characteristic time scales associated 
with elastic stresses. One that characterizes the time it takes for the distorted ring to regain its equilibrium shape, 
$\tau_{N_1}\propto d_{fibre}^2/\left(\Gamma L_1\right)$, and another that is the time it takes for the figure-of-eight defect to untwist, $\tau_{N_2}$. 
We have performed measurements of $\tau_{N_1}$ and $\tau_{N_2}$, and found that, typically, $\tau_{N_2}=3.0\tau_{N_1}$ and that for larger systems  ($d_{drop}\sim 1\mu$m) $\tau_{N_1}=2.0\,m$s. 
We can also estimate the response time of the nematic to the applied $E$ field to scale with $\tau_E\propto \left(\Gamma\epsilon_o\Delta\epsilon E^2\right)^{-1}$. 
Our measured $\tau_E$ confirms the $E^{-2}$ dependence. As $E$ increases $\tau_E$ becomes smaller and the effect of the electric field dominates over the LC elasticity.

\subsection{The effect of AC fields}

In LC displays it is possible to use either DC or AC fields. However, AC fields have a practical advantage over DC ones. AC fields typically produce more stable patterns, lower the threshold voltages and minimize electrochemical deterioration \cite{Kawamoto.2002}. In the case of nematic droplets adsorbed at fibres, and in contact with air, the application of a DC field can disrupt completely the drop, which is pulled towards one of the electrodes. This is due to ionic impurities in the LC compound that aid the formation of space charges. However, the effect of free ions can be significantly minimized by the use of AC fields. 
Here we will not explore the effect of AC {\it vs} DC fields on free ions and how these influence the LC texture, which will be done elsewhere. Instead, in this section we aim to address how AC electric fields affect the nematic textures presented in the previous section.

We again consider that the electric field is applied perpendicular to the fibre. We note that, for both DC and AC fields, if the electric field is oblique to the fibre, this results in a rotation of the defect. For simplicity we assume that the electric field is sinusoidal and is given by Eq.\ref{eqn:Eelec} (with no DC component), such that for frequency $f=0$ we regain the DC limit. Our study reveals that there are three types field strengths $E_{mag}$ to consider: $i)$ below the Freederickzs transition field $E_{mag}<E_c$, $ii)$ just above but close to $E_c$, and $iii)$ $E_{mag}>E_c$. Examples of these cases are show in Fig. \ref{efreq} for different field frequencies $f=0,\,100,\,1000, 10000$ Hz.
For field strengths below $E_c$ (Fig.\ref{efreq}$a_i$), the nematic aligns weakly with the applied field $E$. This can result in a small deformation/contraction of the defect ring. In this situation the response time $\tau_E$ and the restoring time $\tau_{N_1}$ are comparable when $E=E_{mag}$. This means that as the frequency of the field is increased (Fig.\ref{efreq}$a_{ii}$ to \ref{efreq}$a_{iv}$) the elastic stress dominates and the deformation/contraction of the defect is weaker.
If, however, the field strength is strong ($E_{mag}>E_c$) the nematic strongly aligns with the electric field and the defect expands (and twists) to the surface of the drop. In this situation the response time $\tau_E$ is smaller than the restoring time $\tau_{N_2}$. As the frequency is increased the electric field will spend a certain amount of time below $E_c$ which results in an effective response time that is comparable to the restoring time $\tau_{N_2}$. As a consequence the defect shrinks and assumes a shape normally observed for DC fields close to the Freederickzs transition.
The most striking phenomena occurs for fields above the transition but comparable to $E_c$. As the frequency is increased the defect regains its nearly circular shape for $E_{mag}<E_c$, indicating that restoring time $\tau_{N_1}$ is effectively smaller than the response time $\tau_E$, the elastic stress dominates, and the Freederickzs transitions is shifted.

\section{Conclusions}

Nematic droplets suspended on fibres are new LC systems that very much resemble water droplets on cobwebs. They differ from the nematic droplets dispersed in a polymer matrix, 
as the fibre piercing through the droplet confers the droplets a torus-like topology. Here we have studied the equilibrium textures of spherical nematic droplets on fibres under 
several anchoring conditions. We have shown that if a nematic droplet with homeotropic anchoring is pierced by a homeotropic fibre two surface defects are nucleated on antipodal 
positions and at the contact lines, which in principle could be removed by elongating the particle. However, if the droplet has planar degenerate anchoring, and the fibre has either 
homeotropic or planar degenerate anchorings, the nematic can avoid the nucleation of topological defects by twisting along the symmetry axis. Finally, if a homeotropic nematic droplet 
is pierced by a planar fibre a ring defect is nucleated close to the fibre in the centre of the drop. Such a defect may be avoided if the fibre has degenerate planar anchoring. In this case 
the nematic will rotate around the fibre and the orientational field assumes a smooth escaped configuration.

We have also studied the effect of applying an electric field to a homeotropic nematic droplet pierced by a planar fibre. We have started by considering the application of a DC field, perpendicular to the fibre, and have seen that the system undergoes a Freederickzs like transition at a field strength $E_c$ that depends on the size of the system and on the anchoring strength at both the surface of the droplet and of the fibre. Below $E_c$ the defect ring undergoes a small deformation/contraction. Above $E_c$ the ring expands towards the surface of the droplet, similar to what is observer for spherical nematic homeotropic droplets. However, the presence of the fibre forces the ring to twist and assume a figure-of-eight shape.

Finally, we have considered the effect of an AC field. In our model, the frequency of the AC field seems to play a small role. Its effects are more pronounced at field strengths just above the Freederickzs transition, where instead of the figure-of-eight defect spanning throughout the system we observe a ring defect (untwisted) indicating that the AC field shifts the transition to higher field strengths.

We note that in our simulations we did not observe dynamical phenomena similar to that reported in \cite{Geng.2012}. Geng and co-workers applied to the system, modelled in this work, an AC field of fixed frequency $f=50$ Hz. They reported that for field strengths below a critical value the ring defect shifted along the fibre in one direction and on the other a spacial soliton seemed to appear rotating with a frequency that depended on the field strength. We can suggest (at least) two reasons for the discrepancy between the dynamics reported for the real system and the behaviour of our model.

The first is related to the size of the systems used in our simulations, which are two orders of magnitude smaller than the physical systems. The size of the simulated system is limited by the resolution required to describe the defect, of the order of the bulk correlation length, $\xi\sim 15\,n$m, which is orders of magnitude smaller than the droplet radius, in the range $10$ to $30\,\mu$m.

The second reason is related to the presence of ionic impurities in the real systems. Ionic impurities may be present in LCs due to molecular degradation, charge injection, or industrial processes. They may be introduced into the LC matrix during sample assembly through the contamination of substrates or of polymer alignment layers \cite{Ciuchi.2007}. 
In the presence of electric fields free ions will induce flows \cite{Lavrentovich.2010,Lavrentovich.2014} and local distortions of the net electric field, which may dominate the dynamics. Including the effect of ions in our simulations is a major task, well beyond the scope of this work.
 
\section*{Acknowledgements}

We acknowledge the support of the Portuguese Foundation for Science and Technology (FCT) through the grants UID/FIS/00618/2013 and EXCL/FIS-NAN/0083/2012.

\end{document}